\documentclass[aps,prl,preprint,nofootinbib]{revtex4-1}
\usepackage{amsmath,amssymb,graphics,graphicx,color}
\usepackage{hyperref}
\usepackage{slashed}

\begin{document}

\title{The Role of a Heavy Neutrino in the Gamma-Ray Burst GRB221009A}
\author{Kingman Cheung$^{a,b}$}

\affiliation{
  $^a$ Department of Physics and Center for Theory and Computation,
  National Tsing Hua University, Hsinchu 30013, Taiwan\\
  $^b$ Division of Quantum Phases and Devices, School of Physics,
  Konkuk University, Seoul 143-701, Republic of Korea
}
\date{\today}

\begin{abstract}
  Recently, several telescopes, including Swift-BAT, GBM, and LHAASO, 
  have observed the ever highest-energy and long-duration gamma-rays
  from a gamma-ray burst named as GRB221009A (located at a red-shift of
  $z=0.151$) on October 9, 2022.
  Conventional understanding tells us that very high-energy photons produced
  at such a far distance suffer severe attenuation before reaching the Earth.
  We propose the existence of a sub-MeV to $O(10)$ MeV heavy neutrino with a transitional
  magnetic dipole moment, via which the heavy neutrino is produced at the GRB.
  It then travels a long distance to our galaxy and decays into a neutrino and a photon,
  which is observed. In such a way, the original high-energy photon produced
  at the GRB can survive long-distance attenuation.
\end{abstract}

\maketitle

Recently, the ever highest-energy and long-duration
gamma rays were detected on Oct. 9, 2022,
first by the Swift Burst Alert Telescope (BAT) \cite{BAT} 
and the Fermi Gamma-ray Burst Monitor (GBM) \cite{FERMI}.
Subsequently,  observations of the very highest gamma rays were recorded from
the same source by the FERMI-LAT \cite{LAT-1,LAT-2}, LHAASO \cite{LHAASO},
and the Carpet-2 \cite{Carpet-2}. Such a
gamma-ray burst (GRB) was measured at a red-shift of $z=0.151$
\cite{redshift,redshift-2},
which corresponds to about 720 Mpc ($\approx 2  \times 10^{25} \,{\rm m}$).
The energy of the very high energy (VHE) photons is determined to be at least 10 TeV
and perhaps as high as 250 TeV.

The detection of very high-energy photons at such a far distance cannot be explained
by conventional physics.  The current understanding of such VHE photons is due to
the explosion of a supermassive star and the creation of a black hole. In such a violent
environment, it is not difficult to imagine very high-energy collisions taking place and thus
creating a lot of hadrons such as pions, kaons, etc. The energetic neutral pions then decay
into photons and charged pions into muons and neutrinos. However, since the GRB
is at a very far distance from us, the VHE photons will lose most of the energies along
their path to us by pair creation, Compton scattering, and other mechanisms. Thus,
the detection of such VHE photons by satellite experiments or experiments on
the Earth is beyond our understanding.

An explanation of photon-axion-photon conversion
was put forward to explain the anomaly
\cite{Galanti:2022pbg,Baktash:2022gnf,Lin:2022ocj,Nakagawa:2022wwm,Zhang:2022zbm}.
The VHE photons so-produced are converted
into axions by the interaction $ f_a a F_{\mu\nu} \tilde{F}^{\mu\nu}$. The axions then
travel a long distance to near our galaxy without interacting with the intergalactic space
and are converted back to photons in presence of a magnetic field of our galaxy. In this
way, the original energy of the VHE photons is preserved.
Other possible interpretations include Lorentz inverse violation
\cite{Finke:2022swf,Li:2022wxc,Zhu:2022usw},
modification of ultrahigh-energy cosmic ray spectrum
\cite{Das:2022gon,AlvesBatista:2022kpg}.

The idea of photon-axion-photon conversion is based on the fact that axions rarely
interact along the path of propagation.  Another well-known particle that shares this
property is the neutrino.  How does a VHE photon related to a neutrino?
We propose the existence of a heavy neutrino
(denoted by $N$) of a mass around $O(10^{-2}) - O(10)$ MeV,
which has a transitional magnetic dipole moment with an active neutrino. 
In the violent environment around the GRB, it is not difficult to imagine that there
are numerous high-energy hadronic collisions, which produce a large number of
pions, kaons, etc.   Thus, we suggest that the neutral pion can decay into a photon,
an active neutrino, and a heavy neutrino, and also the charged pion into a muon,
a photon, and a heavy neutrino
\begin{equation}
  \pi^0 \to \gamma \gamma^* \to \gamma \nu N \;, \qquad
  \pi^\pm \to \mu^\pm \nu^* \to \mu^\pm \gamma N \;,
\end{equation}
in which the flavor of the active neutrino is not important.  The branching ratio
of this decay is small, but it is not critical in our interpretation, because we do not
know exactly the number of collisions that can happen there.  The more important is
that the lifetime of the heavy neutrino is long enough such that it can survive
the path coming towards our galaxy. We give more detail in the following.
The key ingredient of this interpretation is that once the heavy neutrino $N$ is
produced near or at the GRB, it acquires an energy as high as tens or hundreds of TeV.
It then travels a cosmological distance without decay or attenuation, followed
by its decay $N \to \nu \gamma$ when it comes close to our galaxy.  In this way, we
can observe VHE photons of energies $O(10) - O(100)$ TeV.

The transitional magnetic dipole moment of the heavy neutrino (HN) is parameterized
as
\begin{equation}
  \label{lag}
  {\cal L} = \frac{1}{2} \mu_\nu \overline{\nu_L} \sigma^{\mu\nu} N \, F_{\mu\nu} \;
   + \; {\rm h.c.} \;,
\end{equation}
where $F_{\mu\nu}$ is the field strength of the photon field and $\mu_\nu$ is the
magnetic dipole moment of the active-to-heavy-neutrino transition. This interaction is
responsible for both the production and decay of the HN. Production of $N$ can proceed
in meson decays, such as 
$\pi^\pm \to \mu^\pm \nu^* \to \mu^\pm (\gamma N)$ and
$\pi^0 \to \gamma \gamma^* \to \gamma \nu N$, or Primakoff upscattering
$\nu \gamma^* \to N$ \cite{Magill:2018jla}.

\begin{figure}[t!]
  \centering
  \includegraphics{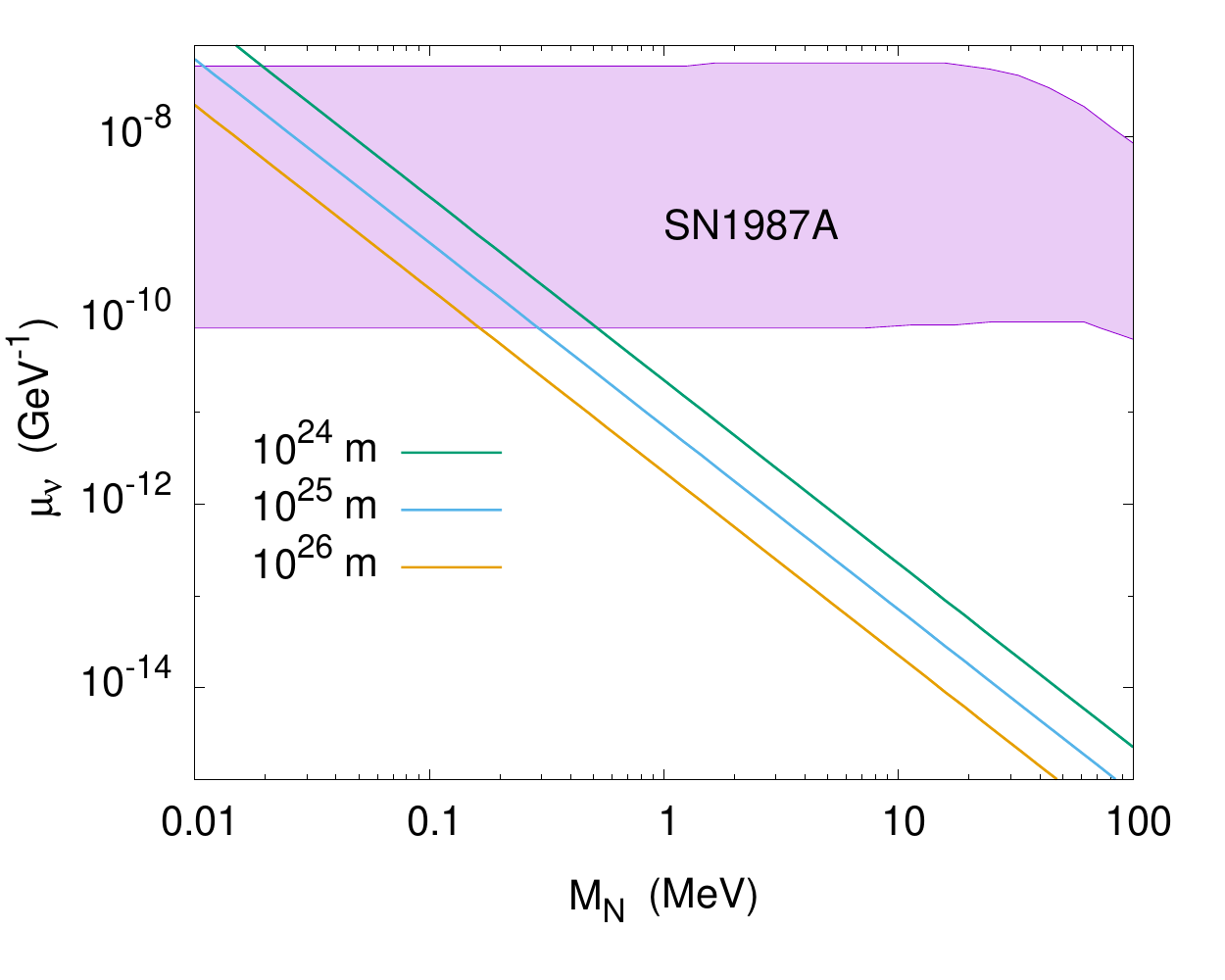}
  \caption{ \label{fig1}
    Contours of constant decay length $L_{\rm decay} = 10^{24}, 10^{25}, 10^{26}$ m
    as a function of $(M_N, \mu_\nu)$.  The  constraint from the supernova SN1987A
    is extracted from Ref.~\cite{Magill:2018jla}. }
\end{figure}

Since there are large uncertainties in the hadronic environment
and energy profile of particles at the GRB, we do not have precise knowledge about 
the number of high-energy pions that can be produced there.
Nevertheless, we know the total amount of energy given off in a typical supernova
explosion is about $10^{44}\, {\rm J} \approx 6.2 \times 10^{62} \, {\rm eV}$
\footnote{ The energy of the GRB221009A was estimated to be about
  $2 \times 10^{54} \, {\rm erg} \equiv 1.2 \times 10^{66}\, {\rm eV} $
  \cite{redshift}.}
and 99\% 
of this energy is carried away by neutrinos. We perform a naive estimate that only
0.1\% of the total energy is going into high-energy hadronic collisions and only 10\%
of this collision energy goes to pions.  The number of
high-energy pions ($~\sim100$ TeV) that can be produced is as many as
$ (10^{58}\, {\rm GeV}) / (10^{14}\, {\rm GeV})  \sim O(10^{44})$. 
Even though the required magnetic dipole moment $\mu_\nu$ is smaller than
$10^{-10} \,{\rm GeV}^{-1}$ (shown later in Fig.~\ref{fig1}) due to the SN1987A
constraint \cite{Magill:2018jla}, the number of $N$ produced should be large enough
to account for the observation.

On the other hand, the decay of $N \to \nu \gamma$ is more important in our
discussion. We calculate the decay width of $N$ based on the
Lagrangian in Eq.~(\ref{lag}). In the rest frame of $N$,
\begin{equation}
  \Gamma ( N \to \gamma \nu) = \frac{| \mu_\nu |^2}{8\pi} M_N^3\;,
\end{equation}
where we have ignored the mass of the active neutrino, which is tiny compared to
$M_N$.  The decay length is then given by
\begin{equation}
L_{\rm decay}  = \beta \gamma c \tau \;,
\end{equation}
where $\beta \simeq 1$, $\gamma \simeq E_N / M_N$ and $\tau = 1 /\Gamma$.
Substituting into the above equation we obtain
\begin{equation}
  L_{\rm decay } = 8 \pi c E_N \frac{1}{M_N^4} \frac{1}{ | \mu_\nu |^2} \;.
\end{equation}
To obtain a representative value of $L_{\rm decay}$ we take $E_N = 100 $ TeV,
$M_N = 10^{-1} $ MeV, and $\mu_\nu = 10^{-9}\,{\rm GeV}^{-1}$, we obtain
\begin{equation}
    \label{final}
  L_{\rm decay} \simeq (5 \times 10^{24}\,{\rm m}) \,
  \left ( \frac{ E_N} { 100 \,{\rm TeV}} \right )\,
  \left( \frac{10^{-1}\, {\rm MeV}}{M_N} \right )^4 \,
  \left ( \frac{ 10^{-9}\,{\rm GeV}^{-1} }{\mu_\nu} \right )^2 \;.
\end{equation}
With this choice of $E_N$, $M_N$ and $\mu_\nu$, the decay length of $N$ falls in the
right ballpark of the distance required to reach from the GRB to our galaxy.
Other choices of $M_N$ and $\mu_\nu$ can be obtained by the scaling in
Eq.~(\ref{final}).

We show a contour plot of decay length $L_{\rm decay}=  10^{24}, 10^{25}, 10^{26}$ m
for the heavy neutrino with $E_N = 100$ TeV 
as a function of $M_N$ and $\mu_\nu$ in Fig. ~\ref{fig1}.
There exist a number of constraints on the transitional magnetic moment of a heavy neutrino
for mass between $10^{-2}$ and 100 MeV \cite{Brdar:2020quo}. Among the constraints,
the relevant one to our study is from supernova SN1987A \cite{Magill:2018jla}.
From the figure, when the magnetic dipole moment is below
$\sim 10^{-10} \,{\rm GeV}^{-1}$
the decay length is in the correct ballpark of the required distance for
$M_N \simeq 0.3 - 10$ MeV. 

To conclude VHE photons produced at the GRB221009A, if traveled
directly to us, would suffer severe attenuation.  We have proposed the
existence of a sub-MeV to $O(10)$ MeV heavy neutrino with a
transitional magnetic dipole moment, via which the heavy neutrino is
produced in the neutral and/or charged pion decays. It then travels a long distance to
our galaxy and decays into a neutrino and a photon, which is
observed. In such a way, the original high-energy photons produced at
the GRB can survive long-distance attenuation.

\begin{acknowledgments}
  We thank D. Alexander Kann for communication on the distance of the GRB.
This work was supported by MoST with grant nos. MoST-110-2112-M-007-017-MY3. 
\end{acknowledgments}

\bibliographystyle{jhep}
\bibliography{references}

\end{document}